\documentclass{article}

\usepackage[utf8]{inputenc} 
\usepackage[T1]{fontenc}    
\usepackage{hyperref}       
\usepackage{url}            
\usepackage{booktabs}       
\usepackage{amsfonts}       
\usepackage{nicefrac}       
\usepackage{microtype}      
\usepackage{amsmath}
\usepackage{xcolor}
\usepackage{soul}
\usepackage{float}
\usepackage{graphicx}

\title{Brown University at TREC Deep Learning 2019}

\author{George Zerveas, Ruochen Zhang, Leila Kim, Carsten Eickhoff\\Brown University, USA\\\texttt{george\_zerveas@brown.edu, carsten\_eickhoff@brown.edu}}

\begin{document}

\maketitle

\begin{abstract}
This paper describes Brown University's submission to the TREC 2019 Deep Learning track. We followed a 2-phase method for producing a ranking of passages for a given input query: In the the first phase, the user's query is expanded by appending 3 queries generated by a transformer model which was trained to rephrase an input query into semantically similar queries. The expanded query can exhibit greater similarity in surface form and vocabulary overlap with the passages of interest and can therefore serve as enriched input to any downstream information retrieval method. In the second phase, we use a BERT-based model pre-trained for language modeling but fine-tuned for query - document relevance prediction to compute relevance scores for a set of 1000 candidate passages per query and subsequently obtain a ranking of passages by sorting them based on the predicted relevance scores.

According to the results published in the official Overview of the TREC Deep Learning Track 2019, our team ranked 3rd in the passage retrieval task (including full ranking and re-ranking), and 2nd when considering only re-ranking submissions.
\end{abstract}

\section{Introduction}
In recent years, deep learning methods have become the standard for solving information retrieval tasks. These methods can effectively map words and phrases to vector representations. These representations can facilitate better matching between phrases that have similar meanings~\cite{word2vec}. Phrases closer in meaning will be represented closer to each other in a vector space. In information retrieval, many ways to develop relevance scores have been used, such as counting word overlap between query and document. Recently, more complex machine learning models use human-verified datasets to train models to assign similarity scores used for rankings. Applying deep learning to Natural Language Processing problems has given rise to new approaches that can better represent a sentence’s meaning using neural networks. For instance, \emph{Long Short Term Memory} models~\cite{lstm} with an attention mechanism allow for word relationships to be constructed between different sentences and thus for words to be better placed in context, rather than just by examining the words closest to them. A breakthrough development in Natural Language Processing, the \emph{BERT} architecture~\cite{bert}, extracts word and consequently sentence representations by masking words throughout a sentence and predicting the omitted words, using self-attention to encode the entire sentence at once. Within the BERT framework, the model can also be trained to predict the next sentence out of a few choices, given an input sentence. \\

Even with these advances, deep learning methods still struggle with some inherent difficulties in IR tasks. These challenges result from discrepancies in query and document vocabulary, limited size of data used for training, and weaknesses in a given human-generated query. In an effort to mitigate these effects, our team’s approach was inspired by an existing method, \emph{doc2query}~\cite{doc2query}, which for a given input document uses a transformer model architecture to predict plausible queries leading to that document. Although it was shown that the expanded documents indeed allowed improved retrieval performance by a downstream ranking model, this approach requires that all documents in the collection of interest are first ``pre-indexed'' by feeding them as input to the transformer model, which is not practical. Instead, we propose a \emph{query2query} method that takes a given query as input and generates several queries similar in meaning. The hope is to create a more powerful query by augmenting the generated queries and the given query into a single representation, which is used to match a desired passage. To complete our architecture, we then feed the expanded queries to a pre-trained BERT model which can predict similarity scores between queries and documents and produce a final ranking. The goal of our approach is to reduce surface form “noise” within a certain query by generating other queries that ask for the same information, but in different ways. By having different representations of the “same” query, we hope to create more holistic queries and as a result obtain an end-to-end method which can generalize better and potentially reduce the problems which modern IR faces.  

\section{Methodology}

\subsection{Neural query expansion}

The data used to train and evaluate our model originated from the publicly available MS MARCO dataset~\cite{ms_marco}. In the frames of the TREC Deep Learning Track competition, the training dataset included 532,761 query - passage pairs labeled as positive for relevance (\texttt{qrels} file). Although the vast majority of passages (97.4\%) are matched only with a single query, for the remaining 2.6\% more than one relevant queries exist (see Table~\ref{tab:num_passages_vs_num_queries}). These are enough to generate 21,582 unique pairs of related queries.

We subsequently regarded the first query of each such pair as a source sentence and the second query as a target sentence for a neural machine translation task. Essentially, this task can be seen as equivalent to paraphrasing an input query into an equivalent query. For this purpose, we trained a transformer model~\cite{transformer} using the OpenNMT~\cite{opennmt} implementation, following a similar pipeline as in~\cite{doc2query}.

To expand the original (source) query, we can then append to the end of the original query the top 3 beams (in terms of estimated log-likelihood) used for the beam search when generating the query2query model's output. The result is an augmented query which consists of 4 approximately equivalent wordings of the same query. Table~\ref{tab:generated_queries} shows several examples were the model rephrased the input query into equivalent formulations.

We found that the quality of the ``equivalent query pairs'' used for training the query2query model is of decisive importance for generating semantically similar queries. Despite its vastly superior size, using a dataset of pairs of queries which yielded the same passages from the \texttt{top1000} data file of the competition (where each query is matched with an unranked set of 1000 potentially relevant candidate passages, and therefore each passage is matched with several queries) often resulted in irrelevant queries being generated by the query2query model, which could easily confound subsequent information retrieval.

\subsection{Re-ranking with BERT}

After expanded queries have been obtained, one can in principle use them as input to any IR method of choice. Due to its proven effectiveness both in other Natural Language Processing tasks as well as document retrieval in particular~\cite{bert_ranking}, we opted for using a BERT model, which has been first pre-trained as part of an unsupervised language modeling objective through input masking (i.e. the original \emph{BERT Large} model, with a hidden state size of 1024).

Because of the significant computational cost of using a BERT model even for inference, one can use as a first step a fast, scalable IR method such as BM25~\cite{bm25} for pre-fetching a limited set of candidate relevant documents from a large collection (e.g. the world wide web). Since we only submitted results for the passage re-ranking task of the competition, this input (an unranked set of the top 1000 potentially relevant passages per query) was already available to us in the \texttt{top1000} data file.

As described in the original BERT paper~\cite{bert}, a pre-trained BERT model can be easily re-purposed to predict an objective of choice by replacing the final (output) layer of the language model with a dense neural layer and a loss function corresponding to the desired objective. The input for this dense layer is the embedding corresponding to the first token (i.e. [CLS]) in the input sequence. In our case, the objective is calculating a relevance score between a given input query (which, as described, has been previously expanded), and a candidate passage or document from the set of top 1000 candidate documents/passages. To obtain such a score, we can simply cast document relevance estimation as a binary classification objective, in which case the score is the estimated (output) probability of relevance. For training the model, the positive (i.e. related) query-passage pairs are contained inside the aforementioned \texttt{qrels} file of ground truth relevance pairs, while negative pairs can be generated by treating any other query-passage pair as unrelated.

Following~\cite{bert_ranking}, we feed the query as sentence A and the passage text as sentence B (using the original notation of~\cite{bert}) after truncating the query to have at most 64 tokens and truncating or padding the passage text such that the concatenation of query, passage and separator tokens have the maximum length of 512 tokens. We fine-tune the model to our re-ranking task using the standard binary cross-entropy loss.

After training the model, one can run it on unseen queries and compute a relevance score for each candidate document/passage, and afterwards simply sort them in order to obtain a final ranking.

\begin{table}[]
\centering
\caption{Number of passages versus the number of queries with which they were matched in the training set ground truth relevance pairings.}
\label{tab:num_passages_vs_num_queries}
\begin{tabular}{l|l}\hline
Num. passages & Num. matched queries \\ \hline
503187 & 1 \\
11328 & 2 \\
1396 & 3 \\
343 & 4 \\
115 & 5 \\
42 & 6 \\
27 & 7 \\
14 & 8 \\
7 & 9 \\
13 & $\geq 10$
 \vspace{6pt}
\end{tabular}
\end{table}

\section{Results}\label{sec:results}

\begin{table}[]
\centering
\caption{Top 3 beam search outputs generated by the query2qury transformer decoder model for an example set of original queries.}
\label{tab:generated_queries}
\resizebox{\textwidth}{!}{%
\begin{tabular}{l|l}\hline
\multicolumn{1}{l|}{\textbf{Original query}} & \textbf{Generated queries} \\ \hline
how long can cooked chicken last in fridge & how long keep refrigerated chicken \\
 & how long will cooked chicken keep \\
 & how long keep chicken \\ \hline
what processes occur during cellular photosynthesis & what is not a waste product of cellular respiration \\
 & what is oxidized during cellular respiration \\
 & what is not a waste product of cellular respiration \\ \hline
what can be done for leg cramps & where do leg cramps usually occur \\
 & what would cause leg cramps \\
 & what would cause cramps \\ \hline
how early can i take a pregnancy test & how soon can a pregnancy test pick up pregnancy \\
 & when to take a pregnancy test \\
 & when can a pregnancy test pick up pregnancy \\ \hline
average salary structural engineer & what is the average salary for a google employee \\
 & what is the average salary for a mechanical engineering \\
 & what is the average starting salary for a mechanical engineering \\ \hline
average tesla cost & what is the cost of the new tesla \\
 & how much money do you save purchasing a tesla \\
 & how much do you have to pay for a tesla
 \vspace{6pt}
\end{tabular}
}
\end{table}

Table~\ref{tab:generated_queries} displays several examples of equivalent queries generated by the query2query transformer decoder. We have observed that the model is able to reliably rephrase the input for queries pertaining to popular subjects, however queries related to more specialized topics or containing exotic terms often yielded semantically unrelated expansions, which could potentially confound subsequent retrieval steps. Moreover, generated expansions sometimes were both grammatically correct as well as aligned with the query thematic, but introduced topical drift, which could at times prove beneficial but potentially also detrimental for downstream retrieval. We believe that a method allowing for better reordering or filtering of the beam output, alongside careful parameter tuning and input pair sampling can improve query expansions in future iterations of the model.

Next, we present the end-to-end performance of our method.

We submitted a single official run for evaluation. Table 3 shows the detailed performance breakdown in terms of Mean Average Precision (MAP), Normalized Discounted Cumulative Gain (nDCG) and Precision at 10 retrieved documents (P@10). The per-topic maximum, median and minimum scores are computed across all 37 submissions and provided by the committee. The categories in the leftmost column indicate the ranges that our topic predictions fall into. For example, in terms of MAP, 2 of our ranking predictions achieve the best score and 29 of them range between the maximum and median scores.\\
\begin{table}[H]
\centering
         \caption{\label{tab:results_comparison} Experiment results of \textit{query2query}}
         \begin{tabular}{c | c | c | c} 
         \hline
         $\text{Number of Topics}$  & $mAP$ & $NDCG$ & $P@10$ \\ [0.5ex] 
         \hline
         $\text{At Best}$ & 2 & 1 & 17  \\ 
         $\text{Best to Median}$ & 29 & 26 & 10 \\ 
         $\text{At Median}$ & 5 & 3 & 13  \\
         $\text{Median to Worst}$ & 6 & 13 & 2  \\
         $\text{At Worst}$ & 1 & 0 & 1  \\
         \hline
        \end{tabular}
        \end{table}

From the results, we observe that 72.1\%, 69.8\% and 93.0\% of our the ranking predictions fall into the median to best region in terms of the three metrics, respectively. It is worth noting that our model achieved top performance for 17 out of 43 test queries in total for the P@10 metrics.

According to the results published in the official Overview of the TREC Deep Learning Track 2019, our team ranked 3rd after \textit{IDST} and \textit{h2oloo} in the passage retrieval task (including full ranking and re-ranking), and 2nd after \textit{IDST} among the re-ranking methods.

\section{Conclusion}\label{sec:conclusion}
This report describes Brown University's entry to the TREC 2019 Deep Learning Track, in which we produced the final ranking of a set of 1000 candidate passages for given queries. Our method aims at enriching the meaning and surface form of a query by expanding it with similar queries, in the hopes that during the subsequent ranking process, the expanded query would provide extra semantic information or vocabulary overlap that would facilitate the retrieval of more relevant documents. \\ 
We found this retrieval method to be promising in terms of retrieval results, albeit with significant margins for future improvement. A natural focus point of future work is improving the semantic similarity between generated queries and the original query. In this work, we simply use the top 3 output beams in terms of estimated log-likelihood. However, different metrics could be used to re-order and prioritize a larger number of generated outputs. In addition, further investigation can be carried out in terms of various ways of synthesizing the query information or condensing the documents' representation.

\section{Acknowledgements}
This research is supported by the Intelligence Advanced Research Projects Activity (IARPA) under grant agreement number IARPA-BAA-18-05.

\bibliographystyle{plain}
\bibliography{main.bib}

\end{document}